# Electromagnetic Field Theory in Superluminal Spacetime

Luca Nanni

luca.nanni@edu.unife.it

**Abstract**

Recently, a number of experimental observations on the superluminal group velocities of pulses propagating in dispersive media have led to reconsidering electromagnetism theory in an unconventional framework. To consider faster-than-light phenomena, it is not necessary to replace the current relativistic theory, but it is sufficient to extend it to superluminal motions in a way that preserves the principle of causality. In the present paper, a new approach to study superluminal motions is proposed, which avoids introducing unphysical complex quantities and allows for the formulation of equations that are covariant according to a hyperbolic metric. In the framework of this formalism, Maxwell equations and the single-photon wave equation are obtained through superluminal transformations of ordinary equations. It is shown that the covariant and contravariant components of the superluminal electromagnetic field determine its magnitude and direction, respectively. Furthermore, the solutions of the transformed Maxwell equations are X-shaped ways propagating in the superluminal spacetime region that is delimited by the infinite light cone and the two-sheet hyperboloid perpendicular to it. Instead, in quantum mechanics, the covariant and contravariant components of the electromagnetic field are proportional to the right- and left-handed helicities of the single photon, respectively.

**1. Introduction**

In recent decades, interest in the study of superluminal phenomena has gradually increased, driven by the results of experiments conducted in various branches of physics [1]. In particle physics, some experiments seem to show a negative square mass of the muon neutrino, which is explained only by its tachyonic behaviour [2-6]. In astrophysics, some observations show apparent superluminal expansions in the core of quasars and microquasars [7-9] or superluminal propagation of photons in the curved space surrounding a black hole [10]. However, in the framework of photonics, the interest in superluminal phenomena has found its most fertile ground, mainly thanks to the pioneering works of Chiao and Nimtz [11-14], which paved the way for designing further and more precise experiments [15-19]. More precisely, the superluminal behaviour of photons is manifested when experiments are carried out using dispersive media, waveguides and devices based on the tunnelling effect. A striking example of superluminal behaviour is propagation with a negative group velocity, where the peak of a pulse propagates across a medium with a negative time delay, appearing as if the peak exited the medium before even entering it.

The majority of physicists have remained sceptical about these results, even if they are not able to refute their veracity with convincing arguments [20-21]. The main problem remains the violation of the principle of causality, with a consequent compromise of the foundations on which the theory of relativity is based. However, in the 1960s, Sudarshan [22-23] - and subsequently Recami [24-25] - proved that it is possible to extend the theory of relativity to superluminal motion without compromising the laws governing it. According to this theory, the speed of light remains an insuperable barrier for both ordinary particles and tachyons. It takes infinite energy to accelerate the bradyon to the speed of light as much as it takes to slow down the tachyon to the same velocity. Furthermore, the theory of extended relativity proposed by Sudarshan and Recami can explain some casual paradoxes that emerge by assuming the existence of tachyons [26-28]. Therefore, it seems that the physical–mathematical apparatus available to describe superluminal phenomena is sufficiently robust and reliable. However, in the framework of photonics, to the best of our knowledge, there is no generalised formulation of Maxwell's equations that is valid, regardless of the relative velocity of the reference frame. More precisely, we refer to the construction of a superluminal spacetime characterised by non-Euclidean geometry, in which the equations of classical and quantum physics maintain the same form as the subluminal equations. This would have the advantage of not introducing unphysical quantities, such as imaginary coordinates or masses, even if this requires changing the spacetime geometry. Moreover, superluminal solutions can be obtained through simple covariant transformations of subluminal solutions. To explain the results of the experiments mentioned above, superluminal solutions of ordinary Maxwell equations have previously been obtained [29-32]. Our goal, however, is to formulate a theory that allows these solutions to be obtained by transforming subluminal solutions through extended Lorentz transformations. The latter has the peculiarity of being continuous in passing from a subluminal to superluminal reference frame, even though this does not imply the possibility of overcoming the light barrier. In our opinion, this approach is simpler than those proposed in the literature and avoids entering speculative fields that do not contribute to scientific progress.

The present paper is organised as follows: In Section 2, a new approach to extending the Lorentz transformations to a superluminal reference frame is discussed. The proposed model does not require the introduction of any unphysical quantities and preserves the form of ordinary relativistic equations using a metric different from the Minkowski one. In Section 3, Maxwell's equations are reformulated in the framework of spacetime extended to superluminal motions. Superluminal transformations of the electromagnetic field components are explicitly provided. In Section 4, the geometry of an electromagnetic pulse in a superluminal frame is investigated. Finally, in Section 5, by exploiting Bialynicki-Birula's work on the formulation of a wave equation for a single photon, the corresponding superluminal version of the wave function is explicitly obtained. The two possible photon helicities are also correlated with the covariant and contravariant components of the spinor.

## 2. Transluminal Lorentz Transformations

Let us consider two reference frames $\mathcal{O}$ and $\mathcal{O}'$ that move at relative velocity $\boldsymbol{u}$ along the $x$ axis. If $u = |\boldsymbol{u}| < c$, then the Lorentz transformations between $\mathcal{O}$ and $\mathcal{O}'$ are as follows:

$$x' = \frac{x - ut}{(1 - u^2/c^2)^{1/2}} \; ; \; y' = y \; ; \; z' = z \; ; \; t' = \frac{t - ux/c^2}{(1 - u^2/c^2)^{1/2}} \tag{1}$$

where the square root of $(1 - u^2/c^2)$ is positive. When $u > c$, the Lorentz factor $(1 - u^2/c^2)^{1/2}$ becomes imaginary, leading to nonphysical coordinates. This problem can be solved by moving it into a complex plane, where the transformations of Eq. (1) take the meaning of $\pi/2$-rotations. Thus, the temporal axis and $x$-axis of the superluminal frame are perpendicular to those of the subluminal frame. However, we want to continue dealing with real quantities without upsetting the formalism of ordinary theory of relativity. To this end, we rewrite Eq. (1) as follows:

$$x' = \frac{x/u - t}{|1/u^2 - 1/c^2|^{1/2}} \; ; \; y' = y \; ; \; z' = z \; ; \; t' = \frac{t/u - x/c^2}{|1/u^2 - 1/c^2|^{1/2}} \tag{2}$$

This formalism allows us to overcome the problem of the imaginary Lorentz factor. In the limit $u \to \infty$, Eq. (2) becomes the following:

$$\lim_{u \to \infty} \frac{x/u - t}{|1/u^2 - 1/c^2|^{1/2}} = -ct \;\; \lim_{u \to \infty} \frac{t/u - x/c^2}{|1/u^2 - 1/c^2|^{1/2}} = -x/c \tag{3}$$

Therefore, in the transformation from the subluminal frame $\mathcal{O}$ to the superluminal $\mathcal{O}'$, the temporal and spatial coordinates are exchanged. This result is not new in the framework of relativistic theories extended to superluminal motions and occurs both in formulations contemplating imaginary quantities [25] and in those that preserve their real values [33]. The exchanged role between spatial and temporal coordinates is extended to both classical quantities and quantum operators. For instance, energy and impulse satisfy the relations $\varepsilon = -p'c$ and $\varepsilon' = -pc$, where $\varepsilon$ and $p$ are the energy and impulse evaluated in $\mathcal{O}$, respectively, and $\varepsilon'$ and $p'$ are the energy and impulse measured in $\mathcal{O}'$ [34-35]. Similarly, Lemke proved that, in quantum mechanics, the subluminal energy operator $i\hbar\partial_t$ is interchangeable with the momentum operator $-i\hbar c\partial_x$, and $i\hbar\partial_t$ is interchanged with $-i\hbar c\partial_x$ [36]. The same occurs in quantum field theory, where the superluminal creation operator is exchanged with the subluminal annihilator and vice versa [37].

Let us return to Eq. (2). The square root appearing in the denominator in the relations giving $x'$ and $t'$ may be positive or negative. It is easy to see that, taking the positive sign, subluminal Lorentz transformations are found. However, it should be noted that the negative sign of the square root is convenient across the light barrier. In fact, because of the relationships in Eq. (3), the choice of the negative sign allows for continuous rotation from the subluminal to the superluminal space. This does not imply that the light barrier can be overcome because, as shown in Eq. (3), it is not defined for $u = c$. This is visualised in the plots in Figure 1.

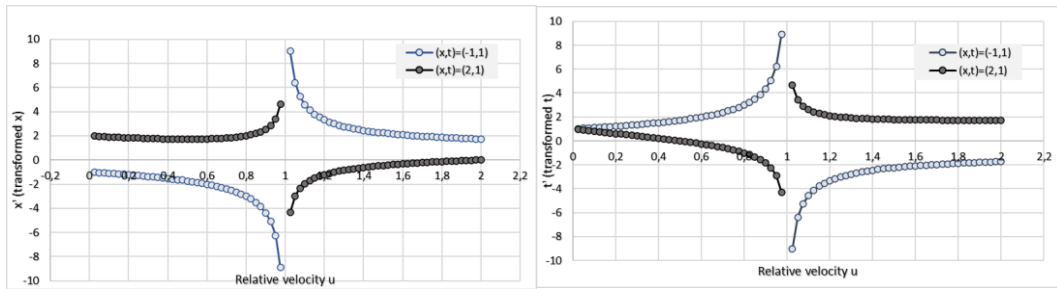

**Figure 1**: Extended subluminal and superluminal transformations for spatial and temporal coordinates concerning the points $(x, t) = (-1,1)$ and $(x, t) = (2,1)$ vs relative velocity $u$

where $c = 1$ was set. As can be seen, the trend of the spatial coordinate versus the velocity $u$ in the subluminal region is the reflected image of the trend of temporal coordinate in the superluminal region, confirming the interchange of the two axes. Thus, the rotation representing transluminal transformations is continuous, except at $u = c$. This is the novelty of our approach compared with other approaches proposed in the literature.

In the Minkowski spacetime, the distance between two events is defined as

$$(ds)^2 = c^2(dt)^2 - (dx)^2 - (dy)^2 - (dz)^2 \tag{4}$$

where the usual metrics $(+, -, -, -)$ are applied. If $(ds)^2 > 0$, then the interval is time-like; otherwise, it is defined as space-like. By applying the transformations in Eq. (2), Eq. (4) can be obtained as follows:

$$(ds')^2 = \gamma'^2 c^2 (dt')^2 (1 - u^2/c^2) + \gamma'^2 (dx')^2 (u^2/c^2 - 1) - (dy')^2 - (dz')^2 \tag{5}$$

If $u > c$ and $\gamma'^2 = |1 - u^2/c^2|^{-1}$, then Eq. (5) becomes the following:

$$(ds')^2 = -c^2(dt')^2 + (dx')^2 - (dy')^2 - (dz')^2 \tag{6}$$

The superluminal interval in Eq. (6) is compatible with the metric $(-, +, -, -)$, where the signs of the temporal and $x$-spatial coordinates are exchanged. The interval $(ds')^2$ may be positive if $(dx')^2 > c^2(dt)'^2 + (dy')^2 + (dz')^2$ or negative if $(dx')^2 < c^2(dt')^2 + (dy')^2 + (dz')^2$. Because our aim is to construct a superluminal spacetime where the equations are consistent with those of the subluminal spacetime, the requirement $(ds')^2 > 0$ must be set. This means limiting the superluminal spacetime to the region in which the inequality $(dx')^2 >$

$c^2(dt')^2 + (dy')^2 + (dz')^2$ holds. Such a region is a four-dimensional vector space in which the covariant components of a generic vector $\boldsymbol{w}$ are $(w_0, w_1, w_2, w_3)$, while the contravariant components are $(-w_0, w_1, -w_2, -w_3)$. We specify that an equation formulated in superluminal spacetime is consistent with its corresponding subluminal counterpart if it is transformed according to the same covariance rule but using a different metric tensor.

In Minkowski's spacetime, once the temporal component has been fixed, the Euclidean subspace can be obtained. In this subspace, the Pythagorean theorem holds:

$$(dr)^2 = (dx)^2 + (dy)^2 + (dz)^2 \tag{7}$$

In superluminal spacetime, the three-dimensional hyperbolic subspace is obtained once the temporal component has been fixed. Eq. (7) then becomes the following:

$$(dr')^2 = (dx')^2 - (dy')^2 - (dz')^2 \tag{8}$$

If $(dr)^2$ is constant, Eq. (7) represents a subluminal ball that, once transformed by the superluminal transformation of Eq. (2), becomes a two-sheet hyperboloid constrained by the inequality $(dx')^2 > (dy')^2 + (dz')^2$. The latter is an infinite cone.

Therefore, in the superluminal space, the wavefront propagates in the region between the infinite cone of light and two-sheeted hyperboloid, here assuming the form of an x. We will return to this in Section 4. Eq. (8) shows that, in the superluminal space, the isotropy condition is violated. However, according to Eq.

(3), for large superluminal velocities $dx' \to -cdt$, whereas $dy' = dy$ and $dz' = dz$. Then, Eq. (8) becomes the following:

$$(dr')^2 = c^2(dt)^2 - (dy)^2 - (dz)^2 \tag{9}$$

which is a hyperboloid obtained by fixing the $x'$-axis instead of the temporal axis. In other words, the three-dimensional space obtained by fixing the $x'$-axis of the superluminal spacetime corresponds to the three-dimensional space obtained by fixing the $t$-axis of the subluminal spacetime. Similarly, if we substitute in Eq. (7) the equalities $dx = ct'$, $dy = dy'$ and $dz = dz'$, the ball equation is obtained as follows:

$$(dr)^2 = c^2(dt')^2 + (dy')^2 + (dz')^2 \tag{10}$$

This shows that superluminal spacetime is also isotropic, just like ordinary spacetime.

The constructed superluminal space has other unexpected physical properties. In this regard, we note that in the reference frame $\mathcal{O}$, the tachyon velocity is always greater than the speed of light. However, in the superluminal frame $\mathcal{O}'$, the tachyon velocity $u = \left|(u_{x'})^2 - (u_{y'})^2 - (u_{z'})^2\right|^{1/2}$ is within the range $0 \leq u < c$ if $(u_{x'})^2 > (u_{y'})^2 + (u_{z'})^2$ and within the range $0 \leq u < \infty$ if $(u_{x'})^2 < (u_{y'})^2 + (u_{z'})^2$. With this in mind, we calculate the invariant $P_\mu P^\mu$, where $P_\mu$ is the superluminal four-impulse $P_\mu = (\varepsilon'/c, p_{x'}, p_{y'}, p_{z'})$:

$$P_\mu P^\mu = m^2 \frac{-c^2 + (u_{x'})^2 - (u_{y'})^2 - (u_{z'})^2}{\left|1 - \frac{(u_{x'})^2 - (u_{y'})^2 - (u_{z'})^2}{c^2}\right|} = -m^2 c^4 \tag{11}$$

With some manipulation returns, it can be the following:

$$P_\mu P^\mu = -\varepsilon'^2 + \left(p_{x'}{}^2 - p_{y'}{}^2 - p_{z'}{}^2\right)c^2 = -m^2 c^4 \tag{12}$$

In Eq. (12), we can recognise the ordinary form of the energy-momentum relation $\varepsilon'^2 = p'^2 c^2 + m^2 c^4$, where $p'^2 = \left(p_{x'}{}^2 - p_{y'}{}^2 - p_{z'}{}^2\right)$. This proves that the tachyon mass is real and positive and that the term $m^2 c^2$ is an invariant under transluminal Lorentz transformations. From Eq. (12), it can be seen that if $(u_{x'})^2 < (u_{y'})^2 + (u_{z'})^2$, then the tachyon energy $\varepsilon'$ is within the range $mc^2 \leq \varepsilon' < \infty$, whereas the impulse $p'$ is within the range $0 \leq p' < \infty$. On the other hand, if $(u_{x'})^2 > (u_{y'})^2 + (u_{z'})^2$ then $0 \leq \varepsilon' < mc^2$ and $0 \leq p' < mc$. Therefore, the latter constraint determines the space of the velocities for which the range of variability of the tachyon quantities is the same as that of the subluminal ones.

Let us define the rest frame for the tachyon. For an ordinary particle, such a frame is one for which the energy is $\varepsilon = mc^2$ and the impulse is $p = 0$. As discussed above, and considering that the superluminal $t'$-axis and $x'$-axis correspond to the subluminal ones exchanged, we expect that for the tachyon rest frame $\varepsilon' = 0$ and $p' = mc$. This is obtained if $u = \infty$ is set as follows:

$$\lim_{u \to \infty} \frac{mc^2}{|1 - u^2/c^2|^{1/2}} = 0 \; ; \; \lim_{u \to \infty} \frac{mu}{|1 - u^2/c^2|^{1/2}} = mc. \tag{13}$$

Eq. (13) holds if $(u_{x'})^2 > (u_{y'})^2 + (u_{z'})^2$, confirming the discussion in Eq. (12). Moreover, because $u$ is greater than the speed of light, the component $(u_{x'})^2$ must be at least equal to $(u_{x'})^2 = c^2 + (u_{y'})^2 + (u_{z'})^2$.

To summarise, the key point of the proposed theory is the change from pseudo-Euclidean geometry to non-Euclidean geometry, which is a necessary condition to avoid obtaining imaginary quantities and preserve the usual notions of covariance and contravariance of tensor calculus. This change is performed by modifying the metric of the Minkowski spacetime.

## 3. Superluminal Maxwell Equations

Through the formalism introduced in Section 2, let us solve the Maxwell equations for a superluminal-charged particle. The easiest approach is to obtain this solution in the tachyon rest frame, where the electric field is static and the magnetic field is zero. As demonstrated in the previous section, the tachyon rest frame is bound by the relation $(x')^2 > (y')^2 + (z')^2$, which is an infinite cone where the temporal and spatial dynamic quantities correspond to the exchanged subluminal quantities. Let us take a closer look at this theoretical aspect. A light pulse emitted by a charged particle in a subluminal frame has a spherical wavefront whose equation is the following:

$$x^2 + y^2 + z^2 = c^2 t^2 \tag{14}$$

By applying the transformations of Eq. (2), the wavefront seen by the superluminal observer is as follows:

$$x'^2 - y'^2 - z'^2 = c^2 t'^2 \tag{15}$$

Because $c^2 t'^2$ must always be positive in the formalism being used, Eq. (15) holds only if $(x')^2 > (y')^2 + (z')^2$, confirming what has been stated above. To obtain the electric field of the charged tachyon in the rest frame, we introduce the tachyonic Klein-Gordon equation [38]:

$$\left(-\hbar^2 \frac{\partial^2}{\partial t'^2} + \hbar^2 c^2 \frac{\partial^2}{\partial x'^2} - \hbar^2 c^2 \frac{\partial^2}{\partial y'^2} - \hbar^2 c^2 \frac{\partial^2}{\partial z'^2}\right)\psi = -m^2 c^2 \psi \tag{16}$$

Eq. (16) has been written using the signature introduced in the previous section. For a virtual photon, one can set $m = 0$. Moreover, because the electric field is static, Eq. (16) becomes the following:

$$\left(\hbar^2 c^2 \frac{\partial^2}{\partial x'^2} - \hbar^2 c^2 \frac{\partial^2}{\partial y'^2} - \hbar^2 c^2 \frac{\partial^2}{\partial z'^2}\right)\psi = 0 \tag{17}$$

which is the superluminal Laplace equation. Because our superluminal formalism must be coherent with the subluminal formalism, we expect the symmetry of the electric field of the charged tachyon to be radial:

$$\lim_{x,y,z \to \infty} E' = 0 \tag{18}$$

Let us solve Eq. (17) by performing the following change of variables:

$$\begin{cases} r' = [(x')^2 - (y')^2 - (z')^2]^{1/2} \\ \theta' = artanh\{[(y')^2 + (z')^2]^{1/2}/x'\}, \\ \varphi' = artan(z'/y') \end{cases} \tag{19}$$

where the constraint $(x')^2 > (y')^2 + (z')^2$ always holds. Using Eq. (19), the superluminal Laplace equation becomes the following:

$$\frac{1}{r'^2}\left[\frac{\partial}{\partial r'}\left(r'^2\frac{\partial}{\partial r'}\right) - \frac{1}{\sinh\theta'}\frac{\partial}{\partial \theta'}\left(\sinh\theta'\frac{\partial}{\partial \theta'}\right) - \frac{1}{\sinh\theta'^2}\frac{\partial^2}{\partial \varphi'^2}\right]\psi = 0 \quad (20)$$

where, to simplify the notation, we set $\hbar = c = 1$. Because we are searching for a solution with radial symmetry, Eq. (20) is reduced to the following:

$$\begin{cases} \frac{1}{r'^2}\left[\frac{\partial}{\partial r'}\left(r'^2\frac{\partial}{\partial r'}\right)\right]\psi = 0 \\ \left(r'^2\frac{\partial}{\partial r'}\right) = -k \end{cases} \quad (21)$$

where $k$ is a numerical constant. Eq. (21) can be solved easily by obtaining:

$$\psi = \frac{k}{r'} + \Lambda \quad (22)$$

where $\Lambda$ denotes the integration constant. As expected, analogous to the subluminal case, $\psi$ represents the potential because of the charged tachyon. Therefore, the modulus of the electric field $\boldsymbol{E'}$ is a radial function:

$$-\frac{\partial \psi}{\partial r'} = \frac{k}{r'^2} \quad (23)$$

However, in the region of superluminal spacetime where we are working, $(r')^2 = (x')^2 - (y')^2 - (z')^2$ and $(x')^2 > (y')^2 + (z')^2$. This means that $r' \to \infty$ even if only $x' \to \infty$, whereas $y'$ and $z'$ are finite. Furthermore, if $x', y', z' \to \infty$, this does not imply that $r' \to \infty$. Therefore, $|\boldsymbol{E'}|$ is not a radial function and does not satisfy our initial requirements. The approach of the rest frame fails and forces us to use a more complicated one in which the model being investigated is solved by applying the superluminal Lorentz transformations of Eq. (2) to the subluminal solution of ordinary Maxwell's equations.

For this purpose, we first introduce the formalism of classical electrodynamics, where the spacetime metric is $(+,-,-,-)$. The electromagnetic tensor $F_{\mu\nu}$ is defined as follows:

$$F_{\mu\nu} = \frac{\partial A_\mu}{\partial x^\nu} - \frac{\partial A_\nu}{\partial x^\mu} \qquad (24)$$

where $A_\mu$ is the four-potential $(U, A_x, A_y, A_z)$. In the subluminal spacetime $A_\mu$ transforms into the following:

$$\tilde{U} = \gamma\left(U - \frac{u}{c}A_x\right), \tilde{A}_x = \gamma\left(A_x - \frac{u}{c}U\right), \tilde{A}_y = A_y, \tilde{A}_z = A_z \qquad (25)$$

The components of the tensor $F_{\mu\nu}$ are related to the electric and magnetic fields by the following relations:

$$\begin{cases} \boldsymbol{H} = \boldsymbol{\nabla} \times \boldsymbol{A} \\ \boldsymbol{E} = -\boldsymbol{\nabla} U - \frac{1}{c}\frac{\partial \boldsymbol{A}}{\partial t} \end{cases} \qquad (26)$$

Using tensor $F_{\mu\nu}$, the Maxwell equations are given by the following:

$$\begin{cases} \frac{\partial F_{\mu\nu}}{\partial x^\sigma} + \frac{\partial F_{\nu\sigma}}{\partial x^\mu} + \frac{\partial F_{\sigma\mu}}{\partial x^\nu} = 0 \\ \frac{\partial F^{\mu\nu}}{\partial x^\nu} = \frac{4\pi}{c} J^\mu \end{cases} \qquad (27)$$

Using Eq. (24) and Eq. (26), we obtain the explicit form of tensor $F_{\mu\nu}$:

$$F_{\mu\nu} = \begin{pmatrix} 0 & \bar{E}_x & \bar{E}_y & \bar{E}_z \\ E_x & 0 & H_z & \bar{H}_y \\ E_y & \bar{H}_z & 0 & H_x \\ E_z & H_y & \bar{H}_x & 0 \end{pmatrix}, \qquad (28)$$

where $\bar{E}_i = -E_i$ and $\bar{H}_i = -H_i$. Substituting Eq. (28) in Eq. (27), Maxwell's equations in differential form are obtained as follows:

$$\begin{cases} \nabla \cdot \boldsymbol{E} = 4\pi\rho \\ \nabla \cdot \boldsymbol{H} = 0 \\ \nabla \times \boldsymbol{E} = -\frac{1}{c}\frac{\partial \boldsymbol{H}}{\partial t} \\ \nabla \times \boldsymbol{H} = \frac{1}{c}\frac{\partial \boldsymbol{E}}{\partial t} + \frac{4\pi \boldsymbol{J}}{c} \end{cases} \qquad (29)$$

These equations are covariant under Lorentz transformations and are equivalent to a system of height differential scalar equations. Notably, this will invoke a generic tensorial transformation:

$$\tilde{F}_{\rho\sigma} = \frac{\partial x^\mu}{\partial x^\rho}\frac{\partial x^\nu}{\partial x^\sigma} F_{\mu\nu} \qquad (30)$$

The following relations are obtained:

$$\begin{cases} \tilde{E}_x = E_x = , \tilde{E}_y = \gamma\left(E_y - \frac{u}{c}H_z\right), \tilde{E}_z = \gamma\left(E_z + \frac{u}{c}H_y\right) \\ \tilde{H}_x = H_x = , \tilde{H}_y = \gamma\left(H_y + \frac{u}{c}E_z\right), \tilde{H}_z = \gamma\left(H_z - \frac{u}{c}E_y\right) \end{cases} \qquad (31)$$

To transform this formalism into a superluminal formalism, Eq. (31) for $u > c$ must be formulated using the metric $(-,+,-,-)$. To simplify the discussion, let us consider the couples $(E_y, H_z)$ and $(E_z, H_y)$, which behave like the couple $(ct, x)$ under the rotations of the ordinary spacetime. Similarly, the couple $(E_x, H_x)$ behaves like $(y, z)$. Using the superluminal transformations of Eq. (2), Eq. (31) becomes the following:

$$\begin{cases} E'_x = E_x = , E'_y = -\gamma'\left(E_y - \frac{u}{c}H_z\right), E'_z = -\gamma'\left(E_z + \frac{u}{c}H_y\right) \\ H'_x = H_x = , H'_y = -\gamma'\left(H_y + \frac{u}{c}E_z\right), H'_z = -\gamma'\left(H_z - \frac{u}{c}E_y\right) \end{cases} \qquad (32)$$

where $\gamma' = \left|1 - (u_{x\prime}^2 - u_{y\prime}^2 - u_{z\prime}^2)/c^2\right|^{1/2}$. The inverse transformations in Eq. (32) are obtained by replacing the positive sign with the negative sign and vice

versa. By applying the latter to Eq. (29), the superluminal Maxwell equations are obtained. Let us initially consider the subluminal four-vector $(c\rho, J)$, whose components are related to the superluminal ones through inverse transformations of Eq. (32).

$$c\rho = \gamma'\left(c\rho' + \frac{u}{c}J'_x\right), J_x = \gamma'\left(J'_x + \frac{u}{c}c\rho'\right), J_y = J'_y, J_z = J'_z \qquad (33)$$

The superluminal counterpart of the first of Eq. (29) is the following:

$$\left(\frac{\partial x'}{\partial x}\frac{\partial}{\partial x'} + \frac{\partial t'}{\partial x}\frac{\partial}{\partial t'}\right)E'_x + \frac{\partial}{\partial y'}\left[\gamma'\left(E'_y + \frac{u}{c}H'_z\right)\right] + \frac{\partial}{\partial z'}\left[\gamma'\left(E'_z - \frac{u}{c}H'_y\right)\right] = \frac{4\pi\gamma'}{c}\left(c\rho' + \frac{u}{c}J'_x\right) \qquad (34)$$

By developing the products and rearranging, one gets the following:

$$\frac{u}{c}\left(-\frac{\partial E'_x}{\partial x'} + \frac{\partial E'_y}{\partial y'} + \frac{\partial E'_z}{\partial z'} - 4\pi\rho'\right) + \left(\frac{\partial H'_z}{\partial y'} - \frac{\partial H'_y}{\partial z'} + \frac{1}{c}\frac{\partial E'_x}{\partial t'} - \frac{4\pi J'_x}{c}\right) = 0 \qquad (35)$$

By combining the first and second equations of Eq. (29) and Eq. (35), we obtain the following:

$$\begin{cases} \left(\dfrac{\partial E'_x}{\partial x'} - \dfrac{\partial E'_y}{\partial y'} - \dfrac{\partial E'_z}{\partial z'}\right) = -4\pi\rho' \\ \left(-\dfrac{\partial H'_z}{\partial y'} + \dfrac{\partial H'_y}{\partial z'}\right) = \dfrac{1}{c}\dfrac{\partial E'_x}{\partial t'} - \dfrac{4\pi J'_x}{c} \end{cases} \qquad (36)$$

By iterating this procedure for all the other six scalar equations that form the system in Eq. (29), the transformation of Maxwell's equations from the subluminal frame to the superluminal frame is complete.

To write the superluminal equivalent of Eq. (29), we introduce the following notation:

$$\begin{cases} E'_v = (E'_x, E'_y, E'_z), E'^v = (E'_x, -E'_y, -E'_z) \\ H'_v = (H'_x, H'_y, H'_z), H'^v = (H'_x, -H'_y, -H'_z) \end{cases} \qquad (37)$$

Using this notation, the superluminal Maxwell equations in differential form are the following:

$$\begin{cases} \boldsymbol{\nabla}' \cdot E'^v = -4\pi\rho' \\ \boldsymbol{\nabla}' \cdot H'^v = 0 \\ \boldsymbol{\nabla}' \times E'_v = \frac{1}{c}\frac{\partial H'^v}{\partial t'} \\ \boldsymbol{\nabla}' \times H'_v = -\frac{1}{c}\frac{\partial E'^v}{\partial t'} + \frac{4\pi J'}{c} \end{cases} \quad (38)$$

where $\boldsymbol{\nabla}' = (\partial/\partial x', \partial/\partial y', \partial/\partial z')$. We define the three-vectors $E'_v$ and $E'^v$ as the covariant and contravariant, respectively. Such vectors do not emerge in a treatment made in the rest frame, which is why the result of Eq. (23) does not satisfy the spherical symmetry requirement. The correct way to write the solution to the field problem generated by a superluminal particle is as follows:

$$E'_v = -\frac{\partial \psi}{\partial x'^v} = -\left(\frac{\partial \psi}{\partial x'}, \frac{\partial \psi}{\partial y'}, \frac{\partial \psi}{\partial z'}\right), E'^v = -\frac{\partial \psi}{\partial x'_v} = \left(-\frac{\partial \psi}{\partial x'}, \frac{\partial \psi}{\partial y'}, \frac{\partial \psi}{\partial z'}\right) \quad (39)$$

By probing the space around the charged tachyon with a particle of unitary charge, Eq. (39) indicates that $E'_v$ is formed by the components of the ordinary potential gradient and represents the magnitude of the force, whereas $E'^v$ gives the direction of the force acting on the particle. To better visualise this result, let us consider a particle constrained to move in the $xy$ plane, along a path parallel to the $y$-axis. Given the hyperbolic geometry of superluminal spacetime, the tachyon accelerates in the opposite direction regarding what occurs in a similar subluminal case. For the example of being considered the proper acceleration, defined as the acceleration that a particle *feels* accelerating from one frame to another is $\alpha = \gamma^3 a$, where $a$ is the three-acceleration $(du/dt)$. By applying superluminal formalism, one obtains

$a' = \gamma'^3 a' = -\left|1 - [u_{x'}{}^2 - u_{y'}{}^2]/c^2\right|^{-3/2}(du/dt')$, which is negative, confirming the above statement.

To conclude this section, we provide the equation for calculating the superluminal equivalent of the four-potential:

$$A'^\mu = -\frac{\partial x'^\mu}{\partial x_\nu} A'^\nu \tag{40}$$

by which the superluminal electromagnetic tensor $F'_{\mu\nu}$ can be easily obtained.

**4. The Geometry of a Superluminal Pulse**

It has been proven that every wave equation, including those of Maxwell, in addition to subluminal solutions, endowed with velocity $c/n$, where $n > 1$ also admits superluminal solutions [29,39-42]. These solutions behave like localised waves that transport electromagnetic energy without dispersion. In the framework of extended relativity, while an electromagnetic pulse manifests as a spherical wave, its superluminal counterpart appears as an X-shaped wave that propagates in a homogeneous medium without distortion [29]. This solution can be obtained by transforming the equation of a subluminal pulse through the transformations of Eq. (2). For this purpose, let us consider a Gaussian pulse in a linear medium with a Gaussian complex coefficient $\delta = \alpha + i\beta$ propagating along the $x$-axis

$$E(\tau) = \frac{1}{2} a_0 e^{-\delta \tau^2} e^{-i\omega_0 \tau} + c.c. \tag{41}$$

where $\tau = (t - x/u_g)$, with $u_g$ group velocity, $a_0$ is the slowly varying temporal envelope $a(t, x)$ at $(t, x) = (0, \mathbf{0})$, and $\omega_0$ is the frequency at the centre of the Gaussian pulse. Rewriting Eq. (41) as

$$E(\tau) = \frac{1}{2} a_0 e^{-(\alpha + i\beta)\tau^2} e^{-i\omega_0 \tau} + c.c. \tag{42}$$

it is clear that the coefficient $\alpha$ determines the pulse duration. We denote the latter a $\tau_p$. Defining $\tau_p$ as the full width at half maximum of $|E(\tau)|^2$, its relationship with coefficient $\alpha$ is given by

$$\tau_p = (2dn/\alpha)^{1/2} \tag{43}$$

where $n$ is the refraction index of the medium and $d$ is its length along $x$-axis. The pulse instantaneous frequency is instead given by the following:

$$\omega_{inst.} = \omega_0 + 2\beta\tau \tag{44}$$

which increases linearly with time.

By applying the superluminal transformation of Eq. (2) to Eq. (42), the transformed time $\tau'$ is obtained:

$$\tau' = t' - \frac{x'}{u'_g} = \gamma' \left[ \left( \frac{x - ut}{u'_g} \right) - \left( t - \frac{ux}{c^2} \right) \right] \tag{45}$$

where $u > c$ and $u'_g$ is the group velocity of the superluminal pulse. The superluminal Lorentz factor $\gamma'$ is given by $|1 - {u'_x}^2/c^2|^{-1/2}$, where $u'_x = \partial x'/\partial t'$. Therefore, the superluminal version of Eq. (42) is the following:

$$E'_v(\tau') = \left( \frac{1}{2} a_0 e^{\alpha \tau'^2} e^{i(\omega_0 \tau' + \beta \tau'^2)}, 0, 0 \right) + c.c \tag{46}$$

Considering that the pulse propagates along the $x$-axis, the velocity $u'_g$ is defined as $\partial \omega'/\partial k'$. The components of the four-vector $k'_\mu$ are $(\omega'/c, k', 0,0)$, whereas those of its contravariant form are $(-\omega'/c, k', 0,0)$. Therefore, the square magnitude of the group velocity is $(u'_g)^2 = [-(\partial \omega'/\partial k')^2 + c^2]$, and the superluminal pulse of Eq. (46) appears as an X-shaped wave travelling in the region delimited by the infinite light cone and two-sheeted hyperboloid, consistent with the constraints $(u_{x'})^2 > (u_{y'})^2 + (u_{z'})^2$ and $(dx')^2 > (dy')^2 + (dz')^2$ discussed in § 2. Notably, the square superluminal electric field is $E'^2 = E'_\nu E'^\nu = (a_0^2 e^{2\alpha \tau'^2}/2)$; this quantity depends on the $u$-cone variable (where $u > c$), which is given by the term $(x - ut)$ through Eq. (45). Because $E'^2$ depends on $x$ only via the variable $(x - ut)$, the pulse propagates without distortion. These pulses are called X-shaped superluminal localised waves and rigidly propagate in a vacuum. The transverse profile of these waves keeps its main peak confined, whereas the lateral peaks expand upon propagation. The obtained result is in full agreement with that obtained by Recami [44]. Therefore, the superluminal transformation of a wave packet satisfying the usual Maxwell equations is an X-shaped wave that naturally satisfies the superluminal Maxwell equations. An observer rigidly fixed to a subluminal frame of reference will see a spherical electromagnetic wave propagating in the region of spacetime confined between the light cone and twofold hyperboloid.

Therefore, it is important to recognise the typical X-shape that the wavefronts seen by the subluminal observer will assume.

The group velocity of the superluminal pulse can be defined as follows:

$$u'_g = |-(\partial \omega'/\partial k')^2 + c^2|^{1/2} \qquad (47)$$

Unlike what occurs for a subluminal pulse, the group velocity is always lower than the speed of light [44-45]. However, in the subluminal sector, tachyonic group velocities leave room for speculation [46-47], whereas in a superluminal frame, the observed velocities can be even lower or greater than $c$, as argued in Section 2. Therefore, the result of Eq. (47) does not represent a contradiction in the theory that we are developing.

In some experiments, light undergoes negative scattering, where phase and energy propagate in a given direction, whereas the group velocity is antiparallel. However, if we analyse the results of these experiments in the superluminal framework, the phase and group velocities are both parallel. Once again, the proposed superluminal formalism restores things by bringing them back to the logic that distinguishes the subluminal world to which we are used.

**5. Superluminal Single-Photon Wave Equation**

In the 1990s, exploiting the surprising similarity between Maxwell's equations and those of relativistic quantum mechanics, Bialynicki-Birula formulated a wave equation for a single photon [48], thus constructing a bridge between classical and quantum electrodynamics. Starting from the Weil equation for the massless neutrino [49], Bialynicki-Birula obtained the following equation:

$$i\hbar \frac{\partial}{\partial t}\psi = [\pmb{\Sigma} \cdot (-i\hbar c \pmb{\nabla})]\psi \tag{48}$$

where $\pmb{\Sigma}$ is a three-vector whose components are the $3 \times 3$ $S_v$ matrices with $v = 1,2,3$, which describe spin-1. In the representation in which the three components of the spin-1 wave function transform under rotations as Cartesian components of a vector, these matrices are given by the following:

$$S_1 = \begin{pmatrix} 0 & 0 & 0 \\ 0 & 0 & -i \\ 0 & i & 0 \end{pmatrix}, S_2 = \begin{pmatrix} 0 & 0 & i \\ 0 & 0 & 0 \\ -i & 0 & 0 \end{pmatrix}, S_3 = \begin{pmatrix} 0 & -i & 0 \\ i & 0 & 0 \\ 0 & 0 & 0 \end{pmatrix} \tag{49}$$

Therefore, the wave function is formed by three components that must be proportional to the vectors $\pmb{E}$ and $\pmb{H}$ of classical electrodynamics. To demonstrate the relationship between Eq. (48) and Maxwell's equations, Bialynicki-Birula rewrites Eq. (48) in vector notation by separating the real and imaginary parts of the wave function. Thus, the following coupled equations are obtained:

$$\begin{cases} \dfrac{\partial}{\partial t}[Re(\psi)] = c\pmb{\nabla} \times [Im(\psi)] \\ \dfrac{\partial}{\partial t}[Im(\psi)] = -c\pmb{\nabla} \times [Re(\psi)] \end{cases} \tag{50}$$

where the identity $[\pmb{\Sigma} \cdot (i\hbar c \pmb{\nabla})] = c\pmb{\nabla} \times \psi$ has been used. Eq. (50) returns Maxwell's equations if the following holds

$$\psi = \frac{1}{\sqrt{2}}\left(\frac{\pmb{E}}{\sqrt{\varepsilon_0}} \pm \frac{\pmb{H}}{\sqrt{\mu_0}}\right) \tag{51}$$

In Eq. (51), $\varepsilon_0$ and $\mu_0$ are the electric permittivity and magnetic permeability in a vacuum, respectively. Eq. (51) ensures that the modulus of the wave function is equal to the electromagnetic energy density. The choice of the positive or negative sign in Eq. (51) determines the two photon helicities. This becomes clear if we consider stationary solutions, for which Eq. (48) becomes the following:

$$[\boldsymbol{\Sigma} \cdot (-i\hbar c \boldsymbol{\nabla})]\psi_\pm = \hbar\omega\psi_\pm \tag{52}$$

Because $\hbar\omega > 0$ (the negative frequency does not add any other information on the dynamics of the photon), depending on the sign $\pm$ in Eq. (51), the spin projection in the direction of motion can be either positive (right-handed helicity) or negative (left-handed helicity). The analogy between Eq. (48) and that of Dirac in chiral form becomes explicit if $\psi$ is written as bispinor $(\psi_+, \psi_-)^t$. In this way, Eq. (48) becomes the following:

$$i\hbar\frac{\partial}{\partial t}\psi = \sigma_3[\boldsymbol{\Sigma} \cdot (i\hbar c \boldsymbol{\nabla})]\psi \tag{53}$$

where $\sigma_3$ is a Pauli-like matrix that exchanges the signs of the bispinor components. This formalism requires that the identity $\psi = \sigma_1 \psi^*$ holds, where $\sigma_1$ is a Pauli-like matrix exchanging the components $\psi_+$ and $\psi_-$. Therefore, the following equality must hold:

$$\psi = (\psi_+, \psi_-)^t = (\psi_+^*, \psi_-^*)^t \tag{54}$$

Let us now transfer the Bialynicki-Birula formalism to the superluminal spacetime with a hyperbolic signature. The square modulus of the transformed wave function must be $|\psi'|^2 = |\boldsymbol{E}'|^2 + |\boldsymbol{H}'|^2$. Through Eq. (37), we obtain the following:

$$|\psi'|^2 = \left| {E'_x}^2 - {E'_y}^2 - {E'_z}^2 + {H'_x}^2 - {H'_y}^2 - {H'_z}^2 \right| \tag{55}$$

To be coherent with the hyperbolic geometry, the constraints $E'_x{}^2 > E'_y{}^2 + E'_z{}^2$ and $H'_x{}^2 > H'_y{}^2 + H'_z{}^2$ must hold. The only way to satisfy Eq. (55) is to write $\psi'$ as follows:

$$\psi'_+ = (E_v + iH_v)\,;\ \psi'_- = (E^v - iH^v) \tag{56}$$

Therefore, the superluminal transform of Eq. (48) is the following:

$$i\hbar \frac{\partial}{\partial t'}\psi'_\pm = [\boldsymbol{\Sigma} \cdot (-i\hbar c \boldsymbol{\nabla}')]\psi'_\pm \tag{57}$$

By separating the real and imaginary components of $\psi'_\pm$, it is possible to write superluminal coupled equations analogous to Eq. (50), which returns the superluminal Maxwell equation given by Eq. (38).

To summarise, the right-handed helicity of superluminal photons is proportional to the covariant components of $\boldsymbol{E'}$ and $\boldsymbol{H'}$, whereas left-handed helicity is proportional to the contravariant components. This allows for writing the superluminal version of Eq. (53):

$$i\hbar \frac{\partial}{\partial t'}\psi' = \sigma'_3[\boldsymbol{\Sigma} \cdot (i\hbar c \boldsymbol{\nabla}')]\psi' \tag{58}$$

where $\sigma'_3$ is a Pauli-like matrix that differs from the matrix $\sigma_3$ of Eq. (53). In fact, if we consider Eq. (58) to be analogous to the chiral form of Dirac's tachyonic equation, then $\sigma'_3$ must be pseudo-Hermitian [50]. In Eq. (58), $\psi' = (\psi'_+, \psi'_-) = \sigma'_1(\psi^\dagger{}_-, \psi^\dagger{}_+)$, where the superscript † transforms the covariant components into contravariant components and vice versa.

Let us return to the right- and left-handed helicity. In classical electrodynamics, the momentum density of radiation is defined as $\boldsymbol{P} = \varepsilon_0(\boldsymbol{E} \times \boldsymbol{H})$. The superluminal version of this quantity splits into covariant $P'_v = \varepsilon_0(E'_v \times H'_v)$ and contravariant $P'^v = \varepsilon_0(E'^v \times H'^v)$ vectors. Therefore, the explicit forms of the photon helicities can be written as follows:

$$\begin{cases} h_+ = \dfrac{S_x(E'_y H'_z - H'_y E'_z) + S_y(E'_z H'_x - H'_z E'_x) + S_z(E'_x H'_y - H'_x E'_y)}{|P'_v|} \\ h_- = \dfrac{S_x(E'_y H'_z - H'_y E'_z) + S_y(E'_x H'_z - H'_x E'_z) + S_z(E'_y H'_x - H'_y E'_x)}{|P'^v|} \end{cases} \quad (59)$$

where $\boldsymbol{S} = (S_x, S_y, S_z)$ is the photon spin vector. Because $|P'_v| = |P'^v|$, from Eq. (59), we can see that the two helicities differ by the opposite sign of the z-components. Therefore, the helicity of the superluminal photon is determined by evaluating the sign of the term $S_z(E'_x H'_y - H'_x E'_y)$. Because helicity is a relativistic invariant, its value must remain so not only under superluminal Lorentz transformations, but also for transluminal transformations.

**6. Superluminal Photon in Cosmology**

The superluminal photon hypothesis has profound implications in contemporary physics. It opens new possibilities for overcoming some unsolved problems in particle physics and cosmology. Notably, vacuum polarisation in quantum electrodynamics (QED) in a background gravitational field induces interactions that violate the strong equivalence principle and affect light propagation. Drummond and Hathrell have shown that this mechanism leads to superluminal photon velocities in the low-frequency limit [51]. Since the discovery by Drummond–Hathrell, superluminal photons have been studied in a variety of curved spacetimes, ranging from Schwarzschild metrics, which describe black holes, to Bondi–Sachs spacetime, which describes gravitational radiation from an isolated source or the Big Bang [52]. One of the fascinating results that has emerged involves the status of the event horizon surrounding a black hole. A photon—or a massive particle—with velocity exceeding the speed of light could escape from within the black hole's horizon. In this case, the location of the effective horizon would become fuzzy on a microscopic scale, with potential consequences for the quantum theory of black holes. In fact, under this hypothesis, the mass spectrum of primordial black holes would be different from what is obtained by applying ordinary gravitational theories and could move experimental research in directions that have not yet been explored. Another fascinating result involves the propagation of photons in the very early universe. Investigations of superluminal photons in the Friedmann-Robertson-Walker (FRW) spacetime have shown that photon velocity increases rapidly at early times, independent of polarisation. Recent works on cosmology in which the fundamental constant $c$ varies over time have shown that an increase in the speed of light in the early universe can resolve the so-called horizon problem, which motivates the inflationary model. Quantitative predictions of the photon velocity values in the strong gravitational fields characterising the inflationary epoch are currently beyond reach. Still, it is intriguing to reflect on how quantum theory predicts that the speed of light increases sharply in the very early evolution of the universe. Therefore, it is reasonable to expect that, in this epoch of perfect symmetry between particles and antiparticles, a considerable quantity of superluminal photons was created. This has relevant implications for the value of the cosmological constant in the early universe because it depends on the speed of light.

The superluminal photon hypothesis could also help explain the nature of dark energy. Recently, Goray has proposed an alternative approach for dark energy based on the photon's illusive mass, which opens up new geometrical perspectives of spacetime [53]. Specifically, photons show a nonzero rest mass interacting with matter that contributes to the universe's curvature. Exploiting this idea, superluminal photons also carry an imaginary mass interacting with strong gravitational fields, further contributing to the spacetime curvature. This suggests a potential link between superluminal photons and dark energy, though such an idea is still far from theory. Furthermore, considering that matter and antimatter were present in equal amounts in the first instants of the Bing Bang, we can infer that luminal and superluminal photons also filled the still limited spacetime in equal measure. This hypothesis would explain why the universe is mainly made of dark energy and the tiny value of its zero-point energy. In fact, it is known from classical physics that the higher the velocity of a tachyon, the lower its energy. In particular, the energy tends to be zero at the limit of infinite velocity. However, studying photon's behaviour at infinite speed must be addressed within the framework of quantum physics; in this sense, much remains to be studied.

The topics introduced in this section are hints at how superluminal photon theory, both classical and quantum, can be applied.

**7. Conclusion**

The current study is mainly motivated by the results of photonic tunnelling experiments, which seem to show unexpected superluminal behaviour of electromagnetic waves when propagating in opaque barriers. However, whenever one talks about superluminal phenomena, the community of physicists is divided between those who a priori consider them as errors of interpretation of the experimental results and those who believe that they are proof that the special theory of relativity is only one of the pieces of a wider mosaic. Recently, Nimtz stated, '*Tunnelling is not a classical process, and as this is not described by the special theory of relativity*' [54. We believe that this statement is more than apt for a phenomenon that, in the quantum framework, is as peculiar as the dual behaviour of matter. Hence, the attempt to construct a relativistic theory that includes both subluminal and superluminal motion and provides an adequate physical–mathematical tool to describe photonic tunnelling. The proposed theory allows for the continuous transformation of a subluminal frame to a superluminal frame by inverting the temporal and spatial axes and changing their signs. Geometrically, this means rotating $\pi/2$ counterclockwise along the temporal axis and $\pi/2$ clockwise along the spatial axis. The signature of the Minkowski spacetime changes, leading to a hyperbolic form of relativistic invariants. This formalism has the advantage of not introducing imaginary quantities and transforming the equations while keeping their structures unchanged. By applying the transluminal transformation, the tachyonic Maxwell equations and photon wave equation can be obtained in a simple manner. By these equations, it is possible to reanalyse the experimental results in a more appropriate context, that is, to explain within a relativistic theory extended to the faster-than-light motions those electromagnetic phenomena that seem to violate the ordinary theory of special relativity. In fact, most studies concerning superluminal photon tunnelling are addressed using the Schrödinger equation, which is neither relativistically invariant nor adequate for describing superluminal phenomena [55-59]. This is because the Schrödinger equation for a massless particle in the presence of a potential barrier is mathematically identical to that of Helmholtz. However, in our opinion, this property does not justify its use in the context being considered. Partha Ghose proposed a theory of superluminal photon tunnelling based on Harish Chandra's equation, which has the advantage of being relativistically invariant, but in any case, it is not formulated for phenomena involving faster-than-light velocities [59]. All that remains is the use of equations that are invariant under superluminal transformations. This has the advantage of providing solutions that would appear to correctly interpret the results of superluminal photon tunnelling. In fact, it has been observed that the simplest solution of superluminal Maxwell equations is precisely the X-shaped wave, which is well suited to a hyperbolic geometry of spacetime and which other authors have already introduced ad hoc to describe superluminal phenomena.

It now remains to address the problem of the principle of causality violation, which could lead to methodological difficulties for a particle, such as a photon. In this regard, let us consider the following one-dimensional example: an observer in a subluminal frame sees an electromagnetic pulse emitted at $(t, x, 0,0)$ and absorbed at $(t', x', 0, 0)$, where $[(x^2 - x'^2) - c^2(t^2 - t'^2)] > 0$. A second observer in a superluminal frame sees the signal absorbed at $(t', x', 0, 0)$ before it is emitted at $(t, x, 0, 0)$. Invoking the Feinberg reinterpretation principle [60], things return to be consistent with the physical reality: *a tachyon sent back in time can always be reinterpreted as a tachyon travelling forward in time because observers cannot distinguish between the emission and absorption of tachyons*. In other words, the second observer sees the pulse emitted at $(t', x', 0, 0)$ and absorbed at $(t, x, 0, 0)$. However, the pulse seen by the second observer is necessarily different from the one seen by the subluminal observer. More precisely, the second observer sees the antiparticle of the particle seen by the first observer. However, it is well known that the photon is its own antiparticle. Therefore, in the framework of the reinterpretation principle, the two observers see exactly the same phenomenon, which solves the problem of causality in the theory formulated in the present work. The next step is to revisit photonic tunnelling within the theory proposed in the present study. This will allow for the re-evaluation of some critical points, such as the Hartman effect and definition of tunnelling time, which have always represented the main object of the dispute regarding superluminal phenomena in optics.